\documentclass[a4paper,11pt]{article}
\usepackage{pos}
\usepackage[most]{tcolorbox}

\newcommand{\be}{\begin{equation}}
\newcommand{\ee}{\end{equation}}

\newcommand{\bea}{\begin{eqnarray}}
\newcommand{\eea}{\end{eqnarray}}
\newcommand{\cPT}{\mathcal{P}\mathcal{T}}
\newcommand{\Herm}{{\rm Herm}}
\newtcbtheorem{conjecture}{Conjecture}{
  enhanced, breakable,
  colback=gray!5, colframe=black,
  fonttitle=\bfseries, separator sign none,
  description delimiters={(}{)}
}{conj}

\title{ERG, axions, and a PT-symmetric accelerating Universe}
\ShortTitle{ERG, Axions and a PT-accelerating Universe}

\author[a,b]{Nick E. Mavromatos}

\affiliation[a]{Theoretical Particle Physics and Cosmology Group, Department of Physics, King's College London, Strand, London WC2R 2LS, UK}

\affiliation[b] {Physics Division, School of Applied Mathematics and Physical Sciences,
 National Technical University of Athens, Zografou Campus, 9 Iroon Polytechneiou Str., Athens 15780, Greece}

\author*[a]{Sarben Sarkar}

\abstract{In conventional quantum mechanics, the Hamiltonian is required to be Hermitian, ensuring real energy eigenvalues. Certain non-Hermitian Hamiltonians  can also possess entirely real spectra when they are invariant under the combined action of parity ($\mathcal P$) and time reversal ($\mathcal T$). Such Hamiltonians are called unbroken parity--time ($\mathcal{PT}$)-symmetric. We conjecture  that,  in a string-inspired scenario,  a repulsive phase of gravity may be due to  a $\mathcal{PT}$-symmetric gravitational theory, characterised by anomalous Chern-Simons terms, coupled to axion fields. The conjecture is based on the behaviour of the chiral anomalous couplings of these axions to electromagnetic potentials. Within the Exact Renormalisation Group (ERG) framework in the modified local potential approximation, we argue that   singularities in the  ERG beta functions for such couplings may indicate an infrared phase transition. At the energy scales of these singularities, we conjecture that the field theory transitions from the Hermitian phase to a  ${\mathcal P}{\mathcal T}$-- symmetric theory, of gravity,  characterised by a repulsive Netwon's ``constant''.  This may induce the acceleration observed in the Universe at late eras, without the need for a dark--energy component.}

\FullConference{13th International Conference on the Exact Renormalization Group (ERG2026)\\
1 - 5 September 2026,  University of Sussex, UK.)}

\begin{document}

\begin{flushleft}
KCL-PH-TH/2026-{\bf 26}   
\end{flushleft}

\maketitle

\section{Introduction and motivation}

The year 1998 witnessed important discoveries, which revolutionized our views on the expanding universe, but also of quantum mechanics. On the cosmological side,  the discovery of distant type Ia supernovae~\cite{SupernovaSearchTeam:1998fmf,SupernovaCosmologyProject:1998vns} (via observations  (with redshift $z \gtrsim 1$))  provided, for the first time, experimental evidence for the accelerated expansion of our universe in modern times. This discovery is confirmed by a plethora of subsequent cosmological observations~\cite{Planck}.
On the quantum mechanical side, the work of~\cite{Bender:1998ke} challenged  our ideas on the role of unitarity and Hermiticity in quantum mechanics:  real energy eigenvalues do not need to correspond only to Hermitian Hamiltonians.  That work  shows that a class of non-Hermitian Hamiltonians, symmetric under P(arity (spatial reflection))T(ime-reversal) (${\mathcal P \mathcal T}$) an antilinear transformation, may also be characterised by real eigenvalues, with potentially both
 fundamental and  physical significance.  This class  extends   to Hamiltonians with other  antilinear symmetries  ~\cite{Mostafazadeh:2003gz,Mannheim:2015hto}. Antilinearity
replaces Hermiticity as an alternative  principle for a viable quantum theory. This  approach suggests many physical tests and applications, ranging from quantum mechanics to quantum field theory (QFT).

Within the QFT framework, a conjecture has been made in Ref.~\cite{Ai:2022csx}, proven to be valid 
for weak-coupling effective theories for a scalar field $\phi$ with quartic interactions at leading order \cite{Chen:2024ynx} in dimensions $D \le 1$; 
\begin{conjecture}{}{analcont}
The Euclidean partition functions of a non-Hermitian $\mathcal{P}\mathcal{T}$-symmetric
$D$-dimensional scalar field theory for a field $\phi$ with interaction $-g\,\phi^4 < 0$, $\mathcal{Z}^{\mathcal{P}\mathcal{T}}(g)$,
and of the Hermitian theory with interaction $\lambda\,\phi^4 > 0$, $\mathcal{Z}^{\rm Herm}(\lambda)$,
are related via
\begin{align}\label{analcont}
\ln \mathcal{Z}^{\mathcal{P}\mathcal{T}}(g)
= \frac{1}{2} \ln \mathcal{Z}^{\rm Herm}\!\left(-g + i\,0^+\right)
+ \frac{1}{2} \ln \mathcal{Z}^{\rm Herm}\!\left(-g - i\,0^+\right).
\end{align}
\end{conjecture}
A natural question is whether the conjecture extends to the generating
functional $\mathcal{Z}[J]$ with source $J$, from which all correlation functions, and hence
renormalized couplings and beta functions, are obtained.
\begin{align}\label{action}
S[\phi;\lambda,J] = \int d^D x \left[\frac12 (\partial\phi)^2
+ \frac12 \phi^2 + \frac{\lambda}{4}\phi^4 - J\phi \right],
\end{align}
with $\lambda = -g$, $g>0$, for the $\mathcal{P}\mathcal{T}$-symmetric theory. For real $J$
the source term is not $\mathcal{P}\mathcal{T}$ invariant; instead, $\mathcal{P}\mathcal{T}$ symmetry implies \footnote{In the $-g\phi^4$ theory the
field is a \emph{pseudoscalar}, $\mathcal{P}:\phi\to-\phi$, while $\mathcal{T}$ leaves
$\phi$ unchanged but, being antilinear, sends $i\to-i$; hence
$\mathcal{PT}:\phi\to-\phi^{*}$ together with complex conjugation of all
c-numbers. }
\begin{align}\label{PTsource}
\mathcal{Z}^{\mathcal{P}\mathcal{T}}[-J] = \left(\mathcal{Z}^{\mathcal{P}\mathcal{T}}[J]\right)^*,
\qquad J\in\mathbb{R},
\end{align}
so that ${\rm Re}\,\mathcal{Z}^{\mathcal{P}\mathcal{T}}[J]$ is even and
${\rm Im}\,\mathcal{Z}^{\mathcal{P}\mathcal{T}}[J]$ is odd in $J$. To see this, write $\mathcal{Z}^{\mathcal{P}\mathcal{T}}[J]=A(J)+i\,B(J)$, with
$A={\rm Re}\,\mathcal{Z}^{\mathcal{P}\mathcal{T}}$ and
$B={\rm Im}\,\mathcal{Z}^{\mathcal{P}\mathcal{T}}$ real functions of the real
source $J$. The two sides of \eqref{PTsource} then read
\begin{equation}
  A(-J)+i\,B(-J) \;=\; A(J)-i\,B(J)\,,
\end{equation}
and equating real and imaginary parts separately gives
\begin{equation}
  A(-J)=A(J)\,,\qquad B(-J)=-B(J)\,,
  \label{eq:ReImparity}
\end{equation}
i.e.\ ${\rm Re}\,\mathcal{Z}^{\mathcal{P}\mathcal{T}}[J]$ is even and
${\rm Im}\,\mathcal{Z}^{\mathcal{P}\mathcal{T}}[J]$ is odd in $J$. Equivalently, on assuming analyticity,
in a Taylor expansion of $\mathcal{Z}^{\mathcal{P}\mathcal{T}}[J]$ in $J$ the coefficients of even powers are real and those
of odd powers are purely imaginary; in particular $B(0)=0$, so the partition
function $\mathcal{Z}^{\mathcal{P}\mathcal{T}}[0]$ is real, as befits a theory
with a real spectrum. 
 
The Hermitian theory, by
contrast, has reflection symmetry $\phi \rightarrow -\phi$, and so 
$\mathcal{Z}^{\rm Herm}[-J;\lambda] = \mathcal{Z}^{\rm Herm}[J;\lambda]$. For $D=0$ the path integral reduces to
$$\mathcal{Z}[J] = \int_{\mathcal C} dx\, e^{-S(x;\lambda,J)}$$ with
$S = \frac12 x^2 + \frac{\lambda}{4}x^4 - Jx$ for a suitable contour $\mathcal C$. For $\lambda=-g<0$ the
integrand decays in four Stokes wedges $W_k$ of opening angle $\pi/4$,
centred at $\arg x = \pi/4,\ 3\pi/4,\ -3\pi/4,\ -\pi/4$ for
$k=1,\dots,4$, respectively. This establishes the exact identity
\begin{align}\label{D0identity}
\mathcal{Z}^{\cPT}[J] + \mathcal{Z}^{\cPT}[-J]
= \mathcal{Z}^{\Herm}[J;-g+i0^+] + \mathcal{Z}^{\Herm}[J;-g-i0^+]\,,
\end{align}
valid for any $J$. In $D=1$ a steepest descent analysis is used involving Euclidean bounces.In $D=1$ the source-generalized conjecture, restricted to the $J$-even part
of $\ln\mathcal{Z}[J]$,
Semiclassically through one-bounce order .
\begin{align}\label{genconj}
\frac12\Big(\ln\mathcal{Z}^{\mathcal{P}\mathcal{T}}[J] + \ln\mathcal{Z}^{\mathcal{P}\mathcal{T}}[-J]\Big)
= {\rm Re}\,\ln\mathcal{Z}^{\rm Herm}[J;-g+i0^+]
+ O\!\left(T\,e^{-2S_b}\right)\,,
\end{align}
where $S_b$ is the bounce action and $T$ is the extent of the Euclidean time
interval, so that $\mathcal{Z}[J]=\mathrm{Tr}\,e^{-TH(J)}$; the error grows
linearly in $T$ because $\ln\mathcal{Z}\simeq -TE_0$ at large $T$, and
(\ref{genconj}) is equivalent to
$E_0^{\mathcal{P}\mathcal{T}}={\rm Re}\,E_0^{\rm Herm}+O(e^{-2S_b})$..Writing $\mathcal{Z}^{\rm Herm}=|\mathcal{Z}^{\rm Herm}|\,e^{i\theta}$ and
$\mathcal{Z}^{\mathcal{P}\mathcal{T}}=|\mathcal{Z}^{\mathcal{P}\mathcal{T}}|\,e^{i\varphi}$,
where $\theta$ (even in $J$) and $\varphi$ (odd in $J$, by~\eqref{eq:ReImparity})
are both $O(e^{-S_b})$, the exact relation~\eqref{D0identity} implies
$\tfrac12\big(\ln\mathcal{Z}^{\mathcal{P}\mathcal{T}}[J]+\ln\mathcal{Z}^{\mathcal{P}\mathcal{T}}[-J]\big)
-{\rm Re}\ln\mathcal{Z}^{\rm Herm}[J;-g+i0^+]
=\ln\cos\theta-\ln\cos\varphi\simeq(\varphi^2-\theta^2)/2=O(e^{-2S_b})$,
so that~\eqref{D0identity} and~\eqref{genconj} are compatible to the order
stated, but cannot both hold exactly\footnote{{In $D=0$ the identity follows from adding integration contours,
which is linear, so it holds for $\mathcal{Z}$ itself rather than for
$\ln\mathcal{Z}$. In $D=1$ a relation for $\mathcal{Z}$ cannot hold, since
$\mathcal{Z}\sim e^{-TE_0}$ at large $T$ and
${\rm Re}\,e^{-TE_0}$ oscillates in sign when ${\rm Im}\,E_0\neq0$; the
logarithmic form, which compares energies, is required. Writing
$\mathcal{Z}^{\rm Herm}=|\mathcal{Z}^{\rm Herm}|\,e^{i\theta}$ with
$\theta=O(e^{-S_b})$, the two forms differ by
$\ln\cos\theta\simeq-\theta^2/2=O(e^{-2S_b})$.}}.

 Inspired by  \eqref{analcont} and \eqref{genconj}  we proposed in Refs.~\cite{Mavromatos:2024ozk, Chen:2025muq}  a specific connection of the observed acceleration of the universe to a non-Hermitian phase at large spatial scales. In such a phase gravity appears effectively repulsive, without invoking dark energy. This connection  may be restricted to  specific types of effective gravitational theory, arising in  string theories~\cite{str}, at energies much lower than the string scale $M_s$ containing  quantum anomalies.

%\section{Anomalies in a (quantum) Gravitational Cosmology}

After string compactification to (3+1)-dimensions, 
a string-inspired cosmology using the running-vacuum-model (RVM) ~\cite{SolaPeracaula:2022hpd, anomalies,Mavromatos:2020kzj}  has been proposed at low-energy (compared to the string energy scale $M_s$). Its primordial, \emph{pre-inflationary phase}, is described by the  effective gravitational action, based on the bosonic massless gravitational string multiplet~\cite{str,kaloper}:
\begin{align}\label{sea3}
S^{\rm eff (I)}_{\rm B} = \int d^{4}x\sqrt{-g}\Big[ -\dfrac{1}{2\kappa^{2}}\, R + \frac{1}{2}\, \partial_\mu b \, \partial^\mu b  + \frac{\sqrt{2} \,\alpha^\prime}{96\, \kappa} \, b(x) \, R_{\mu\nu\rho\sigma}\, \widetilde{R}^{\mu\nu\rho\sigma} + \dots \Big],
\end{align}
where $\alpha^\prime = M_s^{-2}$ is the Regge slope; $\kappa= M_{\rm Pl}^{-1}$ is the inverse of the (3+1)-dimensional reduced Planck mass, with $M_{\rm Pl} = 2.4 \cdot 10^{18}~{\rm GeV}$;  
$\varepsilon_{\mu\nu\rho\sigma}$
is the gravitationally covariant Levi-Civita tensor density; $\nabla_\mu$ the gravitationally covariant derivative, while 
$\widetilde R_{\mu\nu\rho\sigma} = \frac{1}{2} \varepsilon_{\rho\sigma\alpha\beta} \, R_{\mu\nu}^{\quad \alpha\beta}$ is the  dual of the Riemann tensor $R_{\mu\nu\rho\sigma}$, 
 with $\alpha, \beta, \mu, \nu, \rho, \sigma =0, \dots 3$  spacetime indices. The dilaton field is assumed to be stabilised to a constant value \footnote{Throughout, the dilaton is assumed to have been stabilised to a constant value,
$\Phi = \Phi_0$, fixing the string coupling $g_s = e^{\Phi_0}$. This is a
consistent truncation of the four-dimensional string effective action, and it
is required phenomenologically: a massless dilaton would mediate
non-universal long-range forces, in conflict with tests of the equivalence
principle, and would induce a time variation of the gauge couplings, which is
tightly constrained.  In string/D-brane theory, the dilaton can acquire a mass $m_\Phi$ through several mechanisms that generate an effective scalar potential, whose minimum determines the stabilized value of the dilaton. These mechanisms include: string-loop (perturbative) corrections and non-perturbative phenomena (such as gaugino condensation on D-branes, strongly coupled gauge dynamics in the compactified dimensions, and Euclidean D-brane instantons),  
flux compactifications, and supersymmetry breaking in the four-dimensional effective supergravity theory, provided that the Kähler potential and superpotential have an appropriate structure to ensure a stable minimum. 
 The resulting dilaton masses vary widely between mechanisms, flux
stabilisation typically giving the heaviest. We assume that $m_\Phi$ is large
compared with the RG scales relevant to the infrared regime of interest here
(and well above the $\sim 10^{-3}\,{\rm eV}$ fifth-force bound), so that for
$k\ll m_\Phi$ the dilaton can be integrated out, its only remnant being the
fixed value of $g_s$. For $m_\Phi\lesssim k\lesssim M_s$ the dilaton is
dynamical; we neglect its effect on the flow in this UV window.}; $b(x)$ is a massless axion-like particle, which is the (3+1)-dimensional dual~\cite{kaloper} of the spin-one antisymmetric tensor (Kalb-Ramond) field of the string spectrum~\cite{str}. The action \eqref{sea3} is a Chern-Simons (CS) gravity model~\cite{Jackiw:2003pm}, containing gravitational anomalies (associated with the third term of the integrand). As a result of the non-trivial CS interaction, the Einstein equations acquire
the Cotton tensor $C_{\mu\nu}$, which satisfies
$\nabla^\mu C_{\mu\nu}\propto(\partial_\nu b)\,R_{\alpha\beta\gamma\delta}
\widetilde{R}^{\alpha\beta\gamma\delta}$; hence, if ordinary matter
were present, its stress tensor would not be covariantly conserved, in conflict
with the general covariance of the matter action. The pre-inflationary phase
described by \eqref{sea3} must therefore be free of ordinary matter, and the matter
generated at the end of inflation must cancel the gravitational anomaly, as
discussed below. 

Condensation of primordial gravitational-wave (GW) perturbations of the metric field can induce condensates of the CS anomaly in \eqref{sea3}~\cite{Dorlis:2024yqw,Dorlis:2024uei}, which result in an RVM metastable inflationary era~\cite{anomalies,Mavromatos:2020kzj}. At the end of this era, chiral fermionic matter is produced, due to the decay of the unstable vacuum~\cite{anomalies}, which \emph{cancels} the gravitational CS anomaly in \eqref{sea3}. This restores general covariance, as required for a consistent matter quantum field theory in a gravitational background.
Chiral anomalies of the electromagnetic field, however remain, and the gravitational theory has an effective action (related to axion electrodynamics) of the form~\cite{anomalies}:
\begin{align}\label{sea3chiranom}
S^{\rm eff (I)}_{\rm B} = \int d^{4}x\sqrt{-g}\Big[ -\dfrac{1}{2\kappa^{2}}\, R - \frac{1}{4}\,F_{\mu\nu}\, F^{\mu\nu} + \frac{1}{2}\, \partial_\mu b \, \partial^\mu b  -   \frac{\sqrt{2}\,\alpha^\prime}{96\, \kappa} \, b(x) F_{\mu\nu}\, \widetilde{F}^{\mu\nu} + \dots \Big],
\end{align}
where the $\dots$ denote fermions, other matter and radiation fields, of no direct relevance for us.

\section{ERG, ${\mathcal P \mathcal T}$ symmetry and non-Hermitian repulsive Gravity at large scales: towards a reinterpretation of the late-time acceleration of the  Universe}

 Based on the above post-RVM-inflationary scenario \eqref{sea3chiranom}, we study the following axion electrodynamics Lagrangian $\mathcal{L}_{E}$.  We switch to an Euclidean (E) formalism, suitable for a non-perturbative exact renormalization-group (ERG) treatment of the quantum theory~\cite{wetterich1993exact,Rosten:2010vm,Wipf:2013vp} based on the Wetterich
equation for the effective average action $\Gamma_k$~\cite{wetterich1993exact} (denoted often as the functional renormalisation group (FRG)):
\be 
\label{eq:AxEl}
\mathcal{L}_{E}=\frac{1}{4} F_{\mu \nu }F^{\mu \nu }+\frac{1}{2} \partial_{\mu } b\, \partial^{\mu} b + \frac{1}{4} i\mathfrak{g}\, b\, F_{\mu \nu }\, \widetilde{F}^{\mu \nu } \,,
\ee
with gauge fixing $\nabla .{\bf A}=0 \,\rm {and}  \ A_{0}=0$. In the context of our string theory model,  the axion field $a(x)$ corresponds to the massless gravitational axion $b(x)$, and the coupling $\mathfrak{g}$ is given by (\emph{cf.} \eqref{sea3chiranom}):
\begin{align}\label{gvsfb}
\mathfrak{g} =  \sqrt{2} \, \frac{\alpha^\prime}{24\, \kappa}. 
\end{align}
The mass dimension of $\mathfrak{g}$ is $-1$ and so the theory is not perturbatively renormalisable. This theory, however, can be treated approximately using  the Wetterich ERG equation \cite{wetterich1993exact,Rosten:2010vm, Wipf:2013vp}  within the space of couplings restricted to \eqref{eq:AxEl}. Such a non-perturbative FRG treatment has been performed in \cite{Eichhorn:2012uv} and reviewed in \cite{Mavromatos:2024ozk}, where we refer the readers for details. 
The RG beta functions (expressing the running of the renormalized coupling $\mathfrak{g}$ with the RG scale $k$) are not universal, but depend on the gauge fixing scheme adopted~\cite{Eichhorn:2012uv}. This is both a feature of the \emph{non-renormalisability} of the theory in the perturbative sense and a consequence of the truncations encountered in the ERG approach.

 The analysis of \cite{Eichhorn:2012uv}, in flat spacetimes, is used as a guiding formalism that allows the formulation of our conjecture \cite{Mavromatos:2024ozk,Chen:2025muq} in the presence of gravity. 
Using a modified  local potential approximation (LPA) in ERG~\cite{wetterich1993exact, Rosten:2010vm, Dupuis:2020fhh}, the analysis shows the existence of singular points of the non-perturbative RG flow, at which the beta function diverges. These points occur in the \emph{infrared} (IR) (large-distance or low-energy) regime. The  FRG beta function for $\mathfrak g$ reads~\cite{Eichhorn:2012uv}
\begin{align}
\partial_{t} \mathfrak g^{2}=2\mathfrak g^{2}\frac{13\, \mathfrak g^{4}\  -\  8064\  \pi^{2} \mathfrak g^{2}-147456\pi^{4} }{\mathfrak g^{4}-384\pi^{2} \mathfrak g^{2}-147456\pi^{4} }\,, \qquad \partial_t  \equiv k \,\partial_k\,.
\end{align}
 The values of $\mathfrak g$ for which the denominator vanishes are the singular points of the beta function, and have the property~\cite{Mavromatos:2020kzj} that they are \emph{either real or purely imaginary} (coming in  pairs).   
 Due to this singular behavior of the beta function in the Hermitian-theory, it does not appear possible  to connect the interacting theory to the free theory at $\mathfrak g=0$ smoothly via an appropriate RG flow from an UV to an IR fixed point.
 This led us~\cite{Mavromatos:2024ozk} to a new \emph{$\mathcal P\mathcal T$ symmetric interpretation} of the theory, which is conjectural (using conjecture \ref{analcont} and \eqref{genconj}) and crucial to our analysis. The beta function of a non-Hermitan theory is obtained through a ``naive'' analytic continuation of the coupling $\mathfrak g \to i \tilde g$. 
 %Because of this unexpected behaviour, the beta function of a non-Hermitan theory (obtained through a ``naive'' analytic continuation of the coupling $\mathfrak g \to i \mathfrak g$) 
%also shows such singular behavior. 
%However,  there is there a smooth connection between the UV fixed point and the trivial infrared fixed point at zero coupling. 
In this case, too, the beta function exhibits a singularity, located
at $\tilde g_c^2 = \pi^2\big(\sqrt{184320} - 192\big) \simeq 2342.30$, as well
as a non-trivial fixed point at $\tilde g_\star^2 \simeq 175.445$.

In terms of $x\equiv\tilde g^2$ the flow reads
$$\partial_t x = 2x\,(13x^2+8064\pi^2x-147456\pi^4)/(x^2+384\pi^2x-147456\pi^4).$$
For $0<x<\tilde g_\star^2$ one has $\beta(x)>0$, while for
$\tilde g_\star^2<x<\tilde g_c^2$ one has $\beta(x)<0$; since
$\partial_t=k\,\partial_k$, in both cases $x$ is driven towards
$\tilde g_\star^2$ as $k\to\infty$, i.e.\ the fixed point is
ultraviolet (UV) attractive, with $\beta'(\tilde g_\star^2)\simeq-2.16$.
Towards the infrared, trajectories with $x<\tilde g_\star^2$ flow
monotonically to the trivial fixed point $x=0$ (where $\beta\simeq 2x$,
so that $x\propto k^2$), and therefore never encounter the singularity,
which lies on the opposite side of $\tilde g_\star^2$. These trajectories
connect the UV fixed point to the infrared (IR) attractive trivial fixed
point, so that, rather than being asymptotically free, this Abelian theory
is asymptotically safe. 

By contrast, trajectories with
$\tilde g_\star^2<x<\tilde g_c^2$ grow towards the IR and reach the
singularity at $x=\tilde g_c^2$ at a finite scale $k_c$.
%%%%%%%%%%%%%%%
This fixed
point is ultraviolet (UV) attractive: $\beta(\tilde g^2)$ is positive for
$\tilde g^2 < \tilde g_\star^2$ and negative for
$\tilde g_\star^2 < \tilde g^2 < \tilde g_c^2$, so the coupling flows towards
$\tilde g_\star^2$ as the RG scale $k$ increases. In the Hermitian theory the singularity of (10) lies between the trivial
fixed point and the non-trivial one ($\mathfrak{g}^2\simeq6132$ versus
$\mathfrak{g}_\star^2\simeq6298$), and so obstructs any flow connecting them.
In the $\mathcal{P}\mathcal{T}$-symmetric theory the ordering is reversed,
$0<\tilde g_\star^2<\tilde g_c^2$, so that trajectories with
$\tilde g^2<\tilde g_\star^2$ connect the UV fixed point $\tilde g_\star^2$
to the trivial IR fixed point $\tilde g^2=0$ without encountering the
singularity. Thus, rather than being asymptotically free, this Abelian theory
is asymptotically safe. The trivial fixed point $\tilde g^2 = 0$, on the other
hand, is an infrared (IR) attractor.
%%%%%%%%%%%%%%%%%%%%%%%%%
%Inspired by  \eqref{analcont}, we  conjectured in \cite{Mavromatos:2024ozk}  that at such a singular point there is a phase transition and the theory is continued into its non-Hermitian counterpart, which is obtained on writing $\mathfrak g=i\tilde{g} $. In this case, we obtain~\cite{Eichhorn:2012uv}
%\begin{align}
%\partial_{t} \tilde{g}^{2}=\beta_{\tilde{g}}=2\tilde{g}^{2}\frac{13\tilde{g}^{4}\  +\  8064\  \pi^{2} \tilde{g}^{2}-147456\pi^{4} }{\tilde{g}^{4}+384\pi^{2} \tilde{g}^{2}-147456\pi^{4} }\,.
%\end{align}

We make possible inferences for the scale dependence of Newton's constant from the above results, within the cosmological framework of the {\it post-inflationary} string-inspired  effective action \eqref{sea3chiranom}. In \cite{Chen:2025muq}, the analogue of the dimensionless (renormalised (``$R$'')) axion coupling \eqref{gvsfb} 
is: 
\begin{align}\label{axionstringcoupl}
\mathfrak g = \mathfrak{g}_R(k) k = \frac{k\,\mathfrak g}{Z_A(k)Z_b(k)^{1/2}} =
\frac{\sqrt{2}}{24} \, \Big(\frac{M_{\rm Pl, R}(k)}{M_{s}^2}\Big)\, k\,, \,\, 
\eta_A=-\partial_t\ln Z_A,\quad
\eta_b=-\partial_t\ln Z_b.
\end{align}
where $Z_{A,b}$ are
wave-function renormalisations (WFR), and
$k$ is a ``running'' (RG) momentum scale. Eq. \eqref{axionstringcoupl} is  crucial for our conjecture connecting $\cPT$-symmetry and repulsive gravity~\cite{Mavromatos:2024ozk,Chen:2025muq}. When this theory is embedded into a version of \emph{quantum} string-inspired gravity, the effective \emph{low-energy} theory is also  characterized by a renormalised combination of scales $\Big(\frac{M_{\rm Pl}(k)}{M_s(k)}\Big)_R$, running with the scale $k$.

Since $M_s$ is associated with the UV momentum cutoff of the effective low-energy theory, it is reasonable to assume that $M_s$ does \emph{not run} with the RG scale $k$.
However, $M_{\rm Pl}$ runs with $k$: in string theory
$M_{\rm Pl}^2 \propto M_s^2\,(V_6 M_s^6)/g_s^2$, with $V_6$ the volume of the
compact space, so at fixed $g_s$ and $M_s$ the running of $M_{\rm Pl}$
reflects a scale dependence of the effective compactification volume.
Hence the square of the renormalised, dimensionless,  running  axion coupling  \eqref{axionstringcoupl} is $\mathfrak g^2=\frac{1}{288} \frac{1}{\widetilde{\rm G}_{\rm N}} \, \frac{k^2}{M_s^2}\,$ 
where the dimensionless quantity $\widetilde{\rm G}_{\rm N} \equiv {\rm G}_{\rm N} \, M_s^2 $, with $\rm G_N=M_{\rm Pl}^{-2}$ the (3+1)-dimensional gravitational (Netwon's) coupling. In \cite{Mavromatos:2024ozk, Chen:2025muq} it was argued that, in the case of \eqref{sea3}, the RG running of $\widetilde{\rm G}_{\rm N}(k)$ with $k$, 
or equivalently with $t \equiv \ln\Big(\frac{k}{M_s}\Big)$, as follows from \eqref{axionstringcoupl}, is 
such that $\frac{d}{dt}\ln\Big(g^2\,\widetilde{\rm  G}_{\rm N} \Big)= 2 $, implying 
\be g^2 \, \widetilde{\rm  G}_{\rm N} = \mathfrak{D} \exp\Big(2t\Big),\ee
where $\mathfrak{D}$ is a positive integration constant fixed by the boundary conditions of the RG flow. The analysis in \cite{Mavromatos:2024ozk} demonstrates that 
the gravitational constant at the singularity goes to zero.  The finite value $\widetilde{\rm G}_{\rm N} (t=0)$ is identified with Newton's constant in the UV regime.

In \eqref{sea3}, the coupling of the axions to the gauge fields is \emph{metric independent}, due to the topological  nature of the Pontryagin density term. Hence,  the inclusion of quantum graviton fluctuations in the metric, as required for a full QG treatment of the problem,
is {\it not} expected to alter the singularity structure of the axion coupling. This leads to the aforementioned properties of the running gravitational coupling, in particular its vanishing in the infrared and has been demonstrated explicitly in \cite{Chen:2025muq}. There we have used the more complete
Wetterich equation~\cite{wetterich1993exact}, which includes gravitational couplings in a systematic way from the start, to
understand the emergence of this phase. 
The singular structure of the RG flows persists
on including gravitational-couplings, thereby offering further support to the conjecture of \cite{Mavromatos:2024ozk}. 

Within the approximations used in
\cite{Eichhorn:2012uv,Chen:2025muq}, the running of $\mathfrak g$ and $\widetilde G_N$ are encoded in the wave-function factors:
$\partial_t \mathfrak g^2=(2+2\eta_A+\eta_b) \mathfrak g^2$, \, $\partial_t\ln(\mathfrak g^2\widetilde G_{\rm N})=
2+2\eta_A+\eta_b+\eta_N$, $\eta_N=-\partial_t\ln Z_N$ and $Z_N\propto G_{\rm N}^{-1}$ (graviton WFR).
Preservation of the conclusions of \cite{Mavromatos:2024ozk} on the running of the gravitational constant requires 
$2\eta_A+\eta_b+\eta_N=0$ along the RG trajectory. This is not automatic in the truncated coupled system of \cite{Chen:2025muq}, and further work is needed before the running of the gravitational sector in the  sttring-inspired model \eqref{sea3} is understood fully.
Nonetheless, transitioning to the non-Hermitian CS theory, can be achieved irrespective of the above, 
by replacing $\mathfrak g$ with $i\mathfrak g$, which, in the context of \eqref{sea3}, \eqref{eq:AxEl}, \eqref{axionstringcoupl}, is equivalent to:
\begin{equation}\label{kappacomplex}
\kappa \,\, \rightarrow \,\, i \,\tilde \kappa, \quad \tilde \kappa \in \mathbb R\,.
\end{equation}
implying a $\mathcal P\mathcal T$-symmetric non-Hermitian 
action, describing repulsive gravity dynamics:
\begin{align}\label{anti}
  S^{\rm eff \, anti}_{\rm B}&=\; \int d^{4}x\sqrt{-g}\Big[ \dfrac{1}{2\widetilde \kappa^{2}}\, R + \frac{1}{2}\, \partial_\mu b \, \partial^\mu b + i\, \frac{\sqrt{2} \,\alpha^\prime}{96\, \widetilde \kappa} \, b(x) \,   F_{\mu\nu}\, \widetilde{
  F}^{\mu\nu} + \dots \Big],
\end{align}
 As argued in~\cite{Mavromatos:2024ozk}, in this non-Hermitian theory the
running of the gravitational coupling also follows from~\eqref{axionstringcoupl}:
provided $2\eta_A+\eta_b+\eta_N=0$ along the flow,
\begin{equation}
  \widetilde g^{\,2}(t)\,\widetilde{\rm G}_{\rm N}(t)
  =\widetilde g^{\,2}(0)\,\widetilde{\rm G}_{\rm N}(0)\,e^{2t}\,,
  \qquad t=\ln(k/M_s)\,,
  \label{gGrel}
\end{equation}
where $t=0$ is a reference scale in the UV regime, at which
$\widetilde g^{\,2}(0)\lesssim\widetilde g^{\star\,2}$.
As $t\to-\infty$ both sides of~\eqref{gGrel} vanish, since the flow ends at
the trivial IR fixed point, so~\eqref{gGrel} alone does not fix
$\widetilde{\rm G}_{\rm N}$ in the IR. However, near the trivial fixed point
$\beta(\widetilde g^{\,2})\simeq 2\widetilde g^{\,2}$, so that
$\widetilde g^{\,2}(t)\simeq C\,e^{2t}$, with $C$ a constant characterising
the trajectory. Eq.~\eqref{gGrel} then gives
\begin{equation}
  \widetilde{\rm G}_{\rm N}(t\to-\infty)
  =\frac{\widetilde g^{\,2}(0)\,\widetilde{\rm G}_{\rm N}(0)}{C}\,,
\end{equation}
a finite, non-zero value, which we identify with Newton's gravitational
constant at large scales.

%\section{Interpretation in terms of an induced late-time  accelerating  Universe} 

\begin{tcolorbox}[
colback=gray!5,
colframe=black,
title=\textbf{Conjecture 2: ~Non-Hermitcity \& PT-symmetry instead of Dark Energy},
fonttitle=\bfseries,
boxrule=0.8pt,
arc=2mm,
left=2mm,
right=2mm,
top=1mm,
bottom=1mm
]

This scenario may lead to a 
\textbf{reinterpretation of the dark-energy sector of the universe}, provided it survives further formal scrutiny in the gravitational sector. Our conjecture points to the possibility that {\it unbroken} $\mathcal P \mathcal T$ symmetry is a {\it low energy} symmetry of the real world. Repulsive gravity, \emph{not} dark energy, might then  be responsible for the late-time acceleration~\cite{Mavromatos:2024ozk,Chen:2025muq} of the Universe. 
Due to the cancellation (at cosmological scales) of the CS gravitational anomaly during the post-inflationary Universe, 
  the most important  phenomenological consequence for the Universe is the effects of the transition to a $\mathcal P \mathcal T$-symmetric phase at late eras. Such a  transition could leave novel imprints on the associated GW spectra, including contributions to the stochastic GW background. %in analogy, {\it e.g.}, with the effects of phase transitions in the early universe (e.g. QCD, first-order electroweak, topological defects {\it etc.}) on the resultant GW profiles.
%~\cite{Caprini:2015zlo,Caprini:2019egz,Kosowsky:1992rz,Grojean:2006bp,Cutting:2020nla}. 
Understanding cosmological tensions this way could also be an interesting research avenue. In parallel, a more rigorous understanding of the running of Newton's
``constant'' is still required: in particular whether the condition
$2\eta_A+\eta_b+\eta_N=0$ is preserved along the full RG trajectory and how
the RG scale $k$ is to be identified with a physical cosmological scale.
%Such studies are currently in progress.
\end{tcolorbox}

\section*{Acknowledgments}

The work of NEM and SS is supported in part by the Science and Technology Facilities Research Council (UK) under the research grant UKRI4231. 
NEM also acknowledges participation in the COST Association Actions CA21136 “Addressing observational
tensions in cosmology with systematics and fundamental physics (CosmoVerse)” and CA23130 ``Bridging high and
low energies in search of quantum gravity (BridgeQG)”.

\bibliographystyle{apsrev4-2}

\bibliography{PT_talk.bib}

\end{document}